\documentstyle[12pt]{article}
\title{\vbox{\vspace{.1mm}}
 Surface Tension, Surface Stiffness, \\
 and Surface Width \\
 of the 3-dimensional Ising Model on a Cubic Lattice}
\author{ \vbox{\vspace{7mm}}
   {\bf Martin Hasenbusch$^{1}$ and Klaus Pinn$^{2}$} \\[6mm]
$^1\,$Fachbereich Physik, Universit\"at Kaiserslautern,
      \\
      Erwin-Schr\"odinger-Str., D-6750 Kaiserslautern, Germany
      \\
      {\tt \small e-mail kphy55@klio.rhrk.uni-kl.de}
      \\[4mm]
$^2\,$Institut f\"ur Theoretische Physik I, Universit\"at M\"unster,
      \\
      Wilhelm-Klemm-Str.\ 9, D-4400 M\"unster, Germany
      \\
      {\tt \small e-mail pinn@yukawa.uni-muenster.de}     \\[4mm] }
\date{September 1992}

\newcommand{\beff}{\beta_{\rm eff}}
\newcommand{\be}{\begin{equation}}
\newcommand{\ee}{\end{equation}}

\newcommand{\rbo}{\raisebox}
\begin{document}
\maketitle

\vspace{-15cm}
\hfill MS-TPI-92-24
\vspace{15cm}

\maketitle \vfill

\begin{abstract}
 We compute properties of the interface of the 3-dimensional Ising model
 for a wide range of temperatures, covering the whole region
 from the low temperature domain through the roughening transition
 to the bulk critical point.
 The interface tension $\sigma$ is obtained by integrating the
 surface energy density over the inverse temperature $\beta$.
 We use lattices
 of size $L \times L \times T$, with $L$ up to $64$, and $T$ up to $27$.
 The simulations with antiperiodic boundary conditions
 in $T$-direction are done with the Hasenbusch-Meyer interface
 cluster algorithm that turns out to be very efficient.
 We demonstrate that in the rough phase the large
 distance behavior of the interface is well described by
 a massless Gaussian dynamics. The surface stiffness
 coefficient $\kappa$ is determined.
 We also attempt to determine the correlation length $\xi$ and study
 universal quantities like
 $\xi^2 \sigma$ and $\xi^2 \kappa$.
 Results for the interfacial width on lattices up to
 $ 512 \times 512 \times 27$ are also presented.
\end{abstract}

\thispagestyle{empty}

\newpage
\section{Introduction}
 There has been continuous interest in the properties of
 interfaces separating coexisting phases. A prominent role is
 played by the 3-dimensional Ising model that is believed
 to share a universality class with
 binary systems in nature.
 The dominating method for quantitative studies of
 the interface of the 3-dimensional
 Ising model is the Monte Carlo method \cite{Binder}, see e.g.\
 \cite{mon1,mon2,bindersurf,MeyerOrt,MuensKles,BergHans,Sanie}, and
 references cited therein.

 The Ising interface undergoes
 a roughening transition at an inverse temperature
 $\beta_r$ that is nearly twice as large as the bulk
 transition coupling $\beta_c$
 ($= 0.221652(3)$ \cite{betac}). The most precise
 estimate for the roughening coupling is
 $\beta_r = 0.4074(3)$ \cite{martinthesis}.

 For $\beta_c < \beta < \beta_r$ the interface is rough.
 It is believed that its infrared properties can be
 described by a massless Gaussian dynamics.
 This is the basic assumption of the theory of
 capillary waves that is widely used to describe
 long distance properties of rough interfaces.
 A Gaussian behavior in the infrared is also what
 is expected from a Kosterlitz-Thouless
 model in the massless phase \cite{KT}.
 Indeed, it is believed
 that the roughening transition of the 3-dimensional
 Ising model is of the Kosterlitz-Thouless type.
 This belief is strongly substantiated by the renormalization
 group analysis in \cite{martinthesis}.

 At and above the bulk transition temperature the system properties
 become independent of the boundary conditions in the infinite
 volume limit. The interface tension (free energy per unit
 area of the interface) vanishes like
 $\sigma \sim \sigma_0 t^{\mu}$,
 with $t=\vert (\beta - \beta_c) / \beta_c \vert$.
 If Widom scaling holds, the
 exponent $\mu$ should be twice the exponent $\nu$
 that determines the critical behavior of the correlation length.
 The most precise estimates for $\nu$ are in the range
 to $0.624 ... 0.630$ \cite{betac,eps}.
 The amplitude $\sigma_0$ is of particular interest because
 it enters certain universal amplitudes that can also be measured
 in real life systems.

 In this paper, we report on a numerical study of
 properties of the Ising interface
 over a wide range of temperatures: from the low
 temperature regime through the rough domain up
 to the bulk transition region.
 The focus is mainly on the interface free energy,
 the interface tension,
 and on the long distance properties in the rough phase.

 The interface free energy is determined by integrating
 the surface energy over $\beta$. (To the best of our
 knowledge, this method to obtain surface free energies
 was first used by B\"urkner and Stauffer \cite{kner83a}).
 We start the integration both at the
 high and at low temperatures and compare the results.

 Our method allows to include interfaces with extension
 up to $64 \times 64$ in the analysis.
 Close to criticality we also use a
 finite step method that allows to directly obtain the
 change of the surface free energy over a small interval
 $\Delta \beta$.

 From the interface free energies we get estimates for
 the interface tension $\sigma$ and make fits with the
 critical law cited above.
 We also determine the correlation length $\xi$ in order to study the
 quantity $\xi^2 \sigma$ where the factors $t^{-2\nu}$
 and $t^{\mu}$ cancel each other.

 In the rough phase, we study the long distance
 behavior of the interface by measuring block spin
 correlation functions of suitably defined interface
 ``height variables''. An effective coupling $\beff$
 (well known in Kosterlitz-Thouless theory) is
 obtained that parameterizes the asymptotic Gaussian
 dynamics. $\beff$ is related to the surface stiffness
 coefficient $\kappa$ that enters the surface Hamiltonian
 of the capillary wave model.

 We also study the surface width. In the rough phase,
 the width is expected to grow logarithmically with
 the surface extension, with a coefficient that
 is proportional to $\beff$. For a simulation at
 $\beta = \beta_c/0.8$ we verify this behavior with
 good accuracy.

 The paper is organized as follows:
 In section \ref{SECmodel}, we introduce the model and our
 conventions. Interfacial properties and our methods
 to compute them are described in section \ref{SECproper}.
 The cluster algorithm for an efficient simulation of
 the Ising model with antiperiodic boundary conditions
 is explained in detail in section \ref{SECcluster}.
 Details of the data analysis and the
 numerical results are summarized in section \ref{SECresults}.
 Section \ref{SECconclu} contains our conclusions.

\section{The model}\label{SECmodel}

 We consider a simple cubic lattice with extension
 $L$ in $x$- and $y$-direction and with extension $T=2D+1$
 in $z$-direction. The lattice sites $i=(i_x,i_y,i_z)$ have integer
 coordinates, and we adopt the convention that the
 $z$-coordinate runs from $-D$ to $+D$.
 The Ising model is defined by the partition function
\be
  Z = \sum_{ \{ \sigma_i = \pm 1\} } \exp(-\beta H) \, .
\ee
The Hamiltonian $H$ is a sum over nearest neighbor contributions,
\be
  H= - \sum_{<i,j>} k_{ij} \sigma_i \sigma_j  \, .
\ee
 The lattice becomes a torus by defining that
 the uppermost plane is regarded as the lower neighbor plane
 of the lowermost plane. An analog identification is done for
 the other two lattice directions. For the Ising spin field
 $\sigma$ we will use two different boundary conditions:
 Periodic boundary conditions are defined by letting
 $k_{ij}=1$ for all links $< \! i,j \!>$ in the lattice.
 To define antiperiodic boundary conditions in $z$-direction, we
 also set $k_{ij}=1$ with the exception of the
 links connecting the uppermost plane
 ($z=+D$) with the lowermost plane ($z=-D$). These links
 carry an antiferromagnetic factor $k_{ij}=-1$.

\section{Interfacial properties}\label{SECproper}

 In this section, we shall give a short account of important
 interfacial properties: surface width, surface tension
 and surface stiffness.

\subsection{Surface width}\label{SUBSECthick}
 For sufficiently large $\beta$ and large enough $L$, the
 imposure of antiperiodic boundary conditions forces the
 system to develop exactly one interface, a region where
 the magnetization rapidly changes from
 a large negative value to a large positive value.
 (Situations where more than one interface can occur are
 discussed below.) An important property of an interface
 is its width. The definition of the interfacial width is
 not unique. We adopt the following definition:

  A magnetization profile for lattice planes perpendicular to
 the $z$-direction is defined by
\be
 M(i_z) = L^{-2} \sum_{i_x,i_y} \sigma_i \, .
\ee
 First we have to unwrap the torus on the real line.
 One then shifts the interface close to $i_z =0$ by taking
 the following measure. An approximate interface position is defined
 as the value of $i_z$ where the absolute value of the magnetization
 profile takes its minimum. The whole Ising configuration is then
 shifted in $z$-direction such that this minimum comes close to
 $i_z=0$. Note that spins that pass the antiperiodic boundary
 at $i_z=\pm D$ are to be flipped.

 We introduce an auxiliary coordinate $z$ that assumes half-integer
 values (labeling positions between adjacent lattice layers
 perpendicular to the $z$-direction). $z$ takes
 values $-D+1/2,-D+3/2,\dots,D-3/2,D-1/2$. A normalized magnetization
 gradient is introduced via
\be
 \rho (z) = \frac{1}{2\:M_b}
 \left\vert M(z+ \frac12)-M(z-\frac12) \right\vert \, ,
\ee
 where $M_b$ is the positive bulk magnetization, i.e.\
\be
 M_b = (L^2 T)^{-1} \left\vert \sum_i \sigma_i \right\vert \, .
\ee
 For a given configuration of the spin field, the position of
 the interface is defined as the sum over
 $z \rho (z)$. The interface width is defined as the expectation
 value
\be
 W^2 = \bigl\langle \,
 \sum_z \rho(z) \, z^2 -  \bigl( \sum_z \rho(z) \, z \bigr)^2 \,
 \bigr\rangle \, .
\ee
 Especially on small lattices, fluctuations in the two bulk phases
 can deteriorate the results.
 Due to bubbles, $\rho(z)$ can
 be accidentially large even far away from the interface position.
 Since such fluctuations contribute to the interface width
 with a weight proportional to the distance from the interface
 position, the true signal can disappear in the noise.
 One possibility to reduce noise that stems from fluctuations of bubbles
 in the bulk is to take the lattice extension $T$ as small
 as possible. In the framework of a study based on a Metropolis
 algorithm, this approach is proposed in \cite{kner83a,mon90a}.
 However, one has to be careful not to disturb
 the free fluctuation of the interface (the properties of which
 we are interested in). In refs.\ \cite{martinthesis,PRL} it is proposed to
 implement a procedure to remove the bubbles before measurement of the
 actual magnetization profile. Note that when the bulk correlation
 length is small, then also the bubbles are small.
 One then  can assume that the interface width changes little when
 one removes all bubbles.
 The procedure is as follows:
 All nearest neighbor pairs
 with a saturated bond are frozen together. (Note that a bond
 connecting top and bottom layer of the lattice is saturated when
 the spins have different signs.)
 This defines a configuration of clusters. By flipping all spins
 in the largest cluster,
 all first order bubbles (bubbles which do not contain
 smaller bubbles) are completely removed. Iterating the procedure, one
 can quickly get rid of all bubbles from the configuration.
 We denote the interface width (measured
 as described above) on the bubble free configuration by
 $W^2_{0}$.

\subsection{Surface tension}\label{SUBSECtens}

 The surface tension $\sigma$ of a d-dimensional
 Ising model is defined by
\be\label{tensdef}
 \sigma =  \lim_{T \rightarrow \infty} \lim_{L \rightarrow \infty} \,
  \frac {1}{L^{d-1}} (F_I - F_0) \, .
\ee
 Here, $F_I = -\ln Z_I$ is the reduced free energy of the system with
 boundary conditions such
 that an interface perpendicular to the $z$-direction
 is introduced at a fixed position.
 The boundary conditions of the system labeled by the subscript ``$0$''
 are such that no interface is forced into the system.

 There are many possibilities to obtain estimates for
 the surface tension from Monte Carlo simulations on finite lattices,
 see e.g.\ refs.\
 \cite{mon1,mon2,bindersurf,MeyerOrt,MuensKles,BergHans,Sanie}.

 One has to do essentially with two sources of systematic errors
 that are distinct in nature but intimately related.
 Effects from too small $L$ and effects from too small $T$.

 To minimize the finite size effects in $z$-direction we choose
 antiperiodic boundary conditions as described above. These
 boundary conditions do not fix the position of the interface:
 it can wander freely in $z$-direction and is less affected
 by the presence of a boundary compared to a system with fixed
 ``$+-$'' boundary conditions. However, it is still important that
 $T$ is large compared to the width of the interface.

 The finite $L$-effect is as follows: For $\beta_c < \beta < \beta_r$,
 the interface is rough, which means that its thickness grows
 logarithmically with $L$. This means, that if we go to large
 interface areas we simultaneously have to increase $T$ in order to
 avoid strong effects from confining a wildly fluctuating
 surface to a flat box. On the other hand, if we choose
 $L$ too small, tunneling becomes more likely and
 there is a tendency that more than one interface will
 form. Note that in general there will be an odd number
 of interfaces for antiperiodic boundary conditions and
 an even number of interfaces in the case of periodic
 boundary conditions. For very small $L$ the
 notion of an interface can even become meaningless.

 The finite size effects become
 the stronger the closer one approaches
 the bulk critical point where no interface survives the
 thermodynamic limit. This means that close to the bulk
 critical point one needs large and thick lattices to
 get systematic errors under control.

 Let us assume that there is only
 one interface in the system with periodic boundary conditions,
 and that there are no interfaces in the periodic system.
 Then the effect of the free motion of the interface in
 $z$-direction on the interfacial free energy amounts to
 add $\ln T$:
\be \label{frene}
 F_s = F_{a.p.} - F_{p.} + \ln T \, .
\ee
 What happens to the surface free energy if there are
 several interfaces in the system?
 If one assumes that the interfaces do not
 interact which each other, one finds
\be
 \tanh(\exp(-F_s + \ln T)) = \frac{Z_{a.p.}}{Z_{p.}}  \, .
\ee
 If we resolve this equation with respect to $F_s$ we get
\be
  F_s =  \ln T - \ln \bigl( \frac{1}{2} \ln ( \frac {1 + Z_{a.p.}/Z_{p.}}
  {1 - Z_{a.p.}/Z_{p.}}) \bigr) \, .
\ee
 In general on has no direct access to the partition
 function in Monte Carlo simulations.
 (For not too large systems, the surface free energy
 can be obtained directly from a Monte Carlo simulation of
 a statistical ensemble that includes the
 boundary conditions as dynamical variables. These variables
 are updated using a modified cluster algorithm
 \cite{direct}.)

 In this paper we shall employ two methods
 to get estimates for the surface free energy.

\subsubsection{Surface free energy from integration over $\beta$}
\label{SUBSUBSECfreeint}

 Note that the derivative of the free energy with respect to
 the coupling $\beta$ is a well defined observable,
\be
 \frac {\partial F}{\partial \beta} = \langle H \rangle \, .
\ee
 In the case of a single interface one therefore gets
\be
 \frac {\partial F_s}{\partial \beta} = \langle H \rangle_{a.p.}
                                      - \langle H \rangle_{p.} \, .
\ee
 Here, the expectation values are defined in the systems with
 periodic or antiperiodic boundary conditions, respectively.
 Let us introduce the abbreviation
\be
 E_s =  \langle H \rangle_{a.p.} - \langle H \rangle_{p.} \, .
\ee
 The surface free energy can then be obtained by integration over $\beta$:
\be
 F_s(\beta) = F_s( \beta_0) + \int_{\beta_0}^{\beta}
 d \beta' \,  E_s(\beta') \, .
\ee
 Our approach is to compute by Monte Carlo simulation the surface energy
 for $\beta$-values ranging from low temperatures around
 $\beta=0.6$ up to the bulk critical region around $\beta_c$.
 Note that we can integrate our data starting both from
 the hot and the cold side
 since the initial conditions for the integration are
 known in both cases:
 For large $\beta$ we can employ a low temperature expansion
 for the interface tension by Weeks et al.\ (published in
 an article by Shaw and Fisher \cite{ShawFisher}) to obtain
 the surface free energy of an interface {\em at a fixed
 position}.

 In the thermodynamic limit the surface tension vanishes in the high
 temperature phase, while it is finite in the low temperature phase.
 For finite systems the difference $F_{a.p.} - F_{p.}$
 strictly vanishes only at $\beta=0$ where only the entropy and not
 the energy enters the free
 energy. But $F_{a.p.} - F_{p.}$ will remain negligibly small until
 $\beta$ comes close to $\beta_c$. For our numerical purpose we set this
 point where the difference of the energies with periodic and
 antiperiodic boundary conditions exceeds a certain amount
 (which we define as the statistical error we can achieve in
 our simulations). Let us call this coupling $\beta_1$.

 Starting from $\beta_1$ where the difference
 $F_{a.p.} - F_{p.}$ is negligible within the obtainable
 accuracy, we can integrate the energy differences to
 obtain the free energy for any $\beta$. Note that in this
 case we have to take into account the existence of
 more than one interface for a certain range of the
 integration interval. In practice this means that we have to
 take $\ln T$ as the integration constant: $F_s(\beta_1) = \ln T$.

\subsubsection{Surface free energy from finite differences}
\label{SUBSUBSECfreefromfin}

 An alternative way to determine free energies is
 to add finite differences in the
 free energy from small intervals $\Delta \beta$.
 For sufficiently small $\Delta \beta$
 and for not too large transverse lattice extension $L$
 we can get the change of
 the free energy directly from a single Monte Carlo simulation.
 It is easy to show that
\be\label{delF}
 F(\beta + \Delta \beta) =  F(\beta)
  - \ln \langle \exp(-\Delta \beta H) \rangle_{\beta}  \, .
\ee
 We put an extra subscript $\beta$ here to make explicit that the
 expectation value is in the system simulated at inverse temperature
 $\beta$.
 Results based on the usage of eq.\ (\ref{delF}) can be easily checked
 for accuracy and consistency: by simulating at a certain point
 $\beta$ one gets estimates for the free energies in a whole neighborhood of
 $\beta$.
 Note that one can use negative $\Delta \beta$'s as well.
 Now assume that we do another simulation at
 $\beta' > \beta$ not too far away from $\beta$. Then we have two
 sets of estimates
 for the points between $\beta$ and $\beta'$, namely the ones
 from the simulation at $\beta$ with positive $\Delta \beta$'s and the
 ones from the simulation at $\beta'$ with negative $\Delta \beta$'s.
 If all the results are consistent (within the statistical accuracy),
 we assume that the step from $\beta$ to $\beta'$ was safe, and
 we proceed to the next larger $\beta$-value.

\subsubsection{Surface tension from finite $L$ data}

 In this section we shall describe how we extract estimates for
 the surface tension $\sigma$
 from the finite $L$ data for the surface free energy $F_s$.
 The fundamental definition of $\sigma$ as given
 in eq.\ (\ref{tensdef}) requires to actually perform the
 limit $L \rightarrow \infty$. However, we observed that
 with very good precision the surface free energy behaves
 like
\be\label{sigmaL}
 F_s = C_s + \sigma' \, L^2 \,
\ee
 already for moderate surface extension $L$.
 It is therefore natural to identify the coefficient $\sigma'$
 in eq.\ (\ref{sigmaL}) with the surface tension $\sigma$.

 Let us discuss how this definition of
 a surface tension on finite lattices relates to another
 one used in the literature. In \cite{MuensKles}, the
 surface tension is computed via the finite $L$ behavior
 of the energy splitting due to tunneling $E_{0a}$,
\be
 E_{0a} = C \exp(- \sigma L^2) \, .
\ee
 Note that the constant $C$ is not identical with the constant
 $C_s$ introduced in eq.\ (\ref{sigmaL}). However, there
 is an approximate relation between the two constants that
 can be derived by approximating the Ising system in a
 long cylinder by a 1-dimensional Ising model with a $\beta$-value
 chosen according to $2\beta=F_s$. (This approximation
 assumes that the interfaces are sharply defined and
 do not interact with each other. Both conditions are
 fulfilled if $\beta$ and $L$ are large enough.)
 With the abbreviation
 $v=\exp(-2\beta)$ one finds
\be
 E_{0a} = \ln \bigl( (1+v)/(1-v) \bigr) \, .
\ee
 For large $\beta$ one has approximately $E_{0a} \approx 2 v$, and
 thus
\be
 2 v = 2 \exp(- C_s -\sigma L^2) \approx C \exp(-\sigma L^2) \, .
\ee
 So we finally obtain the relation
\be\label{appro}
 \ln 2 - C_s \approx \ln C \, .
\ee
\subsection{Surface stiffness}\label{SUBSECstiff}

 In the theory of interfaces the surface stiffness coefficient
 $\kappa$ plays an important role. It is defined as follows.
 In generalization of the surface tension definition given in eq.\
 (\ref{tensdef}) one defines
\be\label{tensteta}
 \sigma(\theta) = \lim_{T \rightarrow \infty} \lim_{L \rightarrow \infty} \,
 \frac {1}{L^{d-1} \cos(\theta) } (F_I - F_0) \, ,
\ee
 where by suitable boundary conditions in the system ``$I$'' a single
 interface is enforced that makes an angle $\theta$ e.g.\ with the
 $x$-axis.
 Expanding for small inclination angle $\theta$,
\be
 \sigma(\theta) / \cos(\theta) =
 \sigma(0) + \sigma'(\theta) \, \theta
 + \frac12 \, \kappa \, \theta^2 + \dots \, ,
\ee
 one defines the stiffness coefficient $\kappa$,
\be
 \kappa = \sigma(0) + \frac{d^2 \sigma}{d\theta^2} \vert_{\theta=0} \, .
\ee
 This coefficient plays an important role in the capillary wave model
 of rough interfaces \cite{capillary}.
 Roughly speaking, this model assumes that the
 surface dynamics of a rough interface is well described by a
 Gaussian model for surface ``height'' variables $h_(X_1,X_2,...,X_{d-1})$.
 The model Hamiltonian is
\be\label{Hcw}
 H_{cw} =
 \frac12 \int dX_1 dX_2 \dots dX_{d-1} \,
 \sum_{i=1}^{d-1} \kappa_i
 \bigl( \frac{d h}{d X_i} \bigr)^2 \, ,
\ee
 where the $\kappa_i$ are the stiffness coefficients corresponding
 to inclinations of the interface with respect to the $i$'th lattice
 plane perpendicular to the $d$'th direction.
 In our case of a 3-dimensional Ising model on a simple cubic
 lattice, $\kappa_1=\kappa_2 \equiv \kappa$, and we define
\be
 \beff = \frac{1}{\kappa} \, .
\ee
 Capillary wave theory then says that the long distance
 properties should by encoded in a
 $(d-1)$-dimensional Gaussian model (massless free
 field theory) with partition function
\be\label{GaussZ}
 Z_0 = \int \prod_i d h_i \exp  \bigl( - \frac{1}{2\beff}
 \sum_{<i,j>} (h_i - h_j)^2 \bigr) \, .
\ee
 Long distance properties are most systematically studied via
 the block spin renormalization group \cite{Wilson}.
 For the Gaussian model defined through eq.\ (\ref{GaussZ})
 one defines block spins $\Phi_I$ as averages over
 cubic blocks $I$ of size $L_B^{d-1}$:
\be
 \Phi_I = L_B^{-(d-1)} \sum_{i \in I} h_i \, .
\ee
 Usually the renormalization group flow is described in terms
 of effective Hamiltonians parameterized by effective
 coupling constants. For our purpose it is sufficient to
 consider expectation values of block spin observables which
 can be directly measured with the Monte Carlo method.
 We define the two quantities
\be\label{aa1}
 A_{1,l}^{(0)} =
 \langle \frac 1 {l^2} \sum_{<I,J>} (\phi_I-\phi_J)^2 \rangle \, ,
\ee
where $I$ and $J$ are nearest neighbors in the block lattice, and
\be\label{aa2}
 A_{2,l}^{(0)} =
 \langle \frac 1 {l^2} \sum_{[I,K]} (\phi_I-\phi_K)^2 \rangle \, ,
\ee
 where $I$ and $K$ are next to nearest neighbors. $l$ is the
 extension of the block lattice, i.e.\ $l = L/L_B$.
 For the Gaussian model, the $A$'s can be computed exactly
 with the help of Fourier transformation.
 The results for a variety of lattice sizes are quoted in table 10.
 These values are computed for $\beta = 1$. Note
 that the results for arbitrary $\beta'$ can be obtained by just
 multiplying with $\beta'/\beta$.

 How do we now do the blocking for the Ising interface?
 Block spin ``height variables'' $\bar h_I$ are defined as follows:
 Blocks $I$ are defined as sets that are quadratic in
 $x-y$ direction with extension $L_B \times L_B$ and that extend
 through the {\em whole} lattice in $z$-direction. One block thus
 contains $L_B^2 T$ lattice points.
 Now note that such a block can be regarded as if it were
 an Ising system in its own right. A magnetization profile
 and an interface position can be determined exactly as in
 the case of the full lattice. We define
\be
 \bar h_I = \mbox{\ interface position in block \ $I$ \ } \, .
\ee
 Note that the blocked height variables can also be defined
 ``with and without bubbles''.
 Blocked observables for the Ising interface are introduced
 analogously to eqs.\ (\ref{aa1}) and (\ref{aa2}):
\be\label{ia1}
 A_{1,l}^{(\rm Ising)} =
 \langle \frac 1 {l^2} \sum_{<I,J>}
 (\bar h_I-\bar h_J)^2 \rangle \, ,
\ee
where $I$ and $J$ are nearest neighbors in the block lattice, and
\be\label{ia2}
 A_{2,l}^{(\rm Ising)} =
 \langle \frac 1 {l^2} \sum_{[I,K]}
 (\bar h_I-\bar h_K)^2 \rangle \, ,
\ee
where $I$ and $K$ are next to nearest neighbor blocks.
For a rough Ising interface, we define an effective coupling
$\beff$ as follows:
\be
 \beff = \lim_{L_B \rightarrow \infty}
 \frac{ A_{i,l}^{(\rm Ising)} }{ A_{i,l}^{(0)} } \, ,
\ee
 Of course, we expect that the so defined $\beff$ does not depend
 on $i$ or $l$. In Monte Carlo studies we can not really
 do the infinite $L_B$-limit. Instead, one has to convince oneself
 that the results have negligible finite size effects. This can, of course,
 only be checked within the given statistical accuracy.

\section{Cluster updating of Ising interfaces}\label{SECcluster}

 Straightforward application of the bulk cluster
 algorithm \cite{swendsen-wang}
 is not appropriate for a rough interface.
 The interface is correlated on all length scales,
 whereas the bulk correlation
 length is finite. Furthermore, competing interactions
 are induced by the interface, and cluster algorithms
 become inefficient for frustrated systems, see e.g.\
 ref.\ \cite{kandel90a}.
 For the simulations reported on in this paper we used the interface
 cluster algorithm proposed in refs.\ \cite{PRL,martinthesis}
 and a slight modification of it, which is more suitable for
 simulations close to the bulk critical temperature.
 The motivation for these interface cluster algorithms stems from the
 VMR cluster \cite{thegang} algorithm that allows efficient
 updating of 2-dimensional solid-on-solid (SOS) models.

 The basic idea of the algorithm is to map the system which is
 frustrated due to the antiperiodic boundary conditions stochastically
 onto a system without frustration. Then this auxiliary system
 is updated using a standard cluster algorithm \cite{swendsen-wang,wolff},
 possibly with a small modification.

 The update of the auxiliary system fulfills detailed balance. For
 the entire update procedure, detailed balance can be implemented
 by making sure that the a priory probability of selecting
 a specific auxiliary system is not altered when
 the auxiliary system is updated.

 Let us now explain how we get the unfrustrated
 auxiliary Ising systems. The first step is to select
 a symmetry plane of the lattice perpendicular to the $z$-axis.
 This plane can be either put {\em between} two layers of sites
 or {\em onto} a layer of sites. Let us label the first
 alternative by ``B'' (for ``between'')
 and the second by ``O'' (for ``onto'').

 (Strictly speaking, the torus is cut twice by the plane. For
 even $T$ one has two B or two O positions, while
 for odd $T$ there is one B and one O position. The algorithms discussed
 here will essentially make use of only one of these positions, depending
 on where the cluster growth is started.)

 We shall describe below how the
 reflection planes are selected stochastically
 with the right probabilities.

 Let us first consider the case that the plane is put between two
 adjacent layers.
 The plane divides the lattice into two parts $\Lambda_+$ and
 $\Lambda_-$. Note that the points of $\Lambda_-$ can be obtained
 from the points of $\Lambda_+$
 by reflection with respect to the plane,
 denoted by $i \rightarrow r(i)$.

 The Ising variables $s_i$ of the auxiliary system
 are now constructed to describe simultaneous
 flips of the spins $\sigma_{i}$ and $\sigma_{r(i)}$.
 For $i \in \Lambda_+$,
 one substitutes in the original Hamiltonian of the system
\begin{eqnarray}
 \sigma_i & \rightarrow & s_i \sigma_i \nonumber \\
 \sigma_{r(i)} & \rightarrow & s_i \sigma_{r(i)} \, ,
\end{eqnarray}
 and determines the conditional probability distribution of the
 $s$-variables (the $\sigma$'s kept fixed).
 For the partition function of the embedded Ising system one gets
\be\label{baux}
  Z_{\rm aux} =
 \sum_{\{s_i = \pm 1\}} \exp \sum_{\langle i,j \rangle} J_{ij} s_i s_j \, ,
\ee
where the couplings $J_{ij}$ in eq.\ (\ref{baux}) are given by
\be\label{coupeff}
 J_{ij} = \beta ( k_{ij} \sigma_{i}
   \sigma_{j} + k_{r(i)r(j)} \sigma_{r(i)} \sigma_{r(j)}) \, ,
\ee
 and the summation is over links in $\Lambda_+$ only. Note
 that the links connecting
 $\Lambda_+$ and $\Lambda_-$ are not contained in the sum:
 Their energy is invariant under the possible spin flips.

 In the case of the reflection plane being identical with a layer of
 lattice sites
 one has to take into account that the sites on the plane are identical
 with their reflected partners. Hence  the corresponding auxiliary spins
  control only one spin of the original system.
 This leads to a modification of the couplings $J_{mn}$ when both
  sites $m$ and $n$ are part of the reflection plane
\be\label{coupeffo}
 J_{mn} = \beta  k_{mn} \sigma_{m}
   \sigma_{n}  \, ,
\ee
 while couplings connecting sites on the reflection plane with sites on
 the neighbor plane keep the form of eq.\ (\ref{coupeff}).

 In ref.\ \cite{martinthesis} it is
 shown that there is no frustration in the auxiliary system.
 This means that the product of the $J_{ij}$ around an arbitrary closed path
 in $\Lambda_+$ is positive or zero.

 To make sure that the Boltzmann weight of the auxiliary configuration
 is equal to the Boltzmann weight of the original configuration we
 set the spins $s_i = 1 $ before each update.
 After an update of the auxiliary system
 the $\sigma_i$ and $\sigma_{r(i)}$
 are flipped if $s_i = -1$. Otherwise they keep their old value.

 Let us now explain the choice of the reflection plane and the update
 of the auxiliary system.
 In the case of the modified (and simpler) version of the algorithm,
 one selects one of the possible reflection planes
 with uniform probability.
 Then one updates the auxiliary system by generating
 and flipping a single cluster using the delete probabilities
\be\label{pdelSW}
 p_{{\rm del},ij} =
 \left\{ \begin{array}{ll}
         \exp(- J_{ij} (1 + s_i s_j) & \mbox{if $ J_{ij} >  0$}
         \nonumber  \, , \\
         \exp(+ J_{ij} (1 - s_i s_j) & \mbox{if $ J_{ij} \leq 0$} \, .
         \end{array}
 \right.
\ee

 Since we want to update interface properties we start the cluster
 at a randomly chosen site on the reflection
 plane in the case of the O position and at a site next to the
 reflection plane in the case of the B position.

 Let us label this version of the algorithm by ``S'' (for ``simple'').
 This version is suitable
 when the width of the interface is of the order of the thickness
 of the lattice $T$.

 In the case of a well defined interface with a width small
 compared to $T$ a more sophisticated procedure for the
 choice of the reflection plane is necessary.
 The corresponding version of the algorithm will be labeled
 with a ``C'' (for ``cut'').
 As in the case of the VMR algorithm for SOS models, the
 crucial point is that the plane cuts the interface, thus
 dividing it into valleys and mountains which are candidate
 objects to be flipped.

 Again we shall describe the procedure first for reflection
 planes located between two adjacent layers of the lattice.
 For a given configuration of the Ising spins $\sigma$, we
 consider the set of all unsaturated bonds that point in
 $z$-direction.
 Let us denote this set by $\cal{B}$.
 If there is a well defined interface, then obviously
 a large portion of $\cal{B}$ will be contained in the interface.
 The following prescription will therefore
 lead to a reasonable frequency of reflection planes that
 cut the interface: With uniform probability select
 a bond from the set $\cal{B}$ and take as reflection plane
 the plane that cuts this bond.

 We have to demonstrate that this choice respects the
 detailed balance condition: The first observation is
 that the selected bond will stay in $\cal{B}$ since
 its value cannot be changed by cluster flips.
 However, we have to make sure that this same bond is
 selected with exactly the same probability when the
 next cluster building is prepared. This can only
 be guaranteed if the size of $\cal{B}$, i.e.\ the
 number of unsaturated links stays unchanged during the
 update. This can be achieved by using the following
 modified delete probabilities for bonds in $z$-direction
 (for the other bonds the
 usual Swendsen-Wang choice
 defined in eq.\ (\ref{pdelSW}) stays in power)
\be\label{ccdd}
 p_{{\rm del},ij} =
 \left\{ \begin{array}{ll}
         1 & \mbox{if $J_{ij} = 0$} \nonumber \, , \\
         0 & \mbox{if $J_{ij} \neq 0$} \, .
         \end{array}
 \right.
\ee

 In order to properly select a reflection plane coinciding with
 one of the lattice layers, one proceeds as described above but
 afterwards shifts the reflection plane by $\pm 1/2$,
 selecting each of the alternatives with probability $1/2$,
 respectively.

 As in the case of the S algorithm we perform a single cluster
 update. We also  start the cluster
 at a randomly choosen site on the reflection
 plane in the case of the O position and at a site next to the
 reflection plane in the case of the B position.

 To restore ergodicity, which is not satisfied by the deleting
 probabilities defined in
 eq.\ (\ref{ccdd}), we perform one Metropolis sweep
 after each update.

 To summarize, the Ising cluster algorithm comes in four
 brands that can be labeled by
 BS,OS,BC,OC. As was the case for the VMR algorithm
 for the SOS models, a mixture of two algorithms of type
 B and O is necessary
 to defeat critical slowing down. The S version performs better
 close to bulk critical point, and the C version is to be
 preferred if the interface is well defined.

 Let us close this section with the remark that
 one can show that the algorithm
 described in this section becomes identical with the VMR algorithm
 in the limit of infinite couplings in $z$-direction, i.e.\ in
 the SOS limit. The BC version becomes the I-algorithm while the OC
 version corresponds to the H-algorithm of ref.\ \cite{thegang}.

\section{Data analysis and Monte Carlo results}\label{SECresults}

 In this section we discuss in some detail how we
 evaluated our Monte Carlo data and summarize our results.

\subsection{Surface free energies and surface tension}
\label{SUBSECmcfe}

 As described in section \ref{SUBSUBSECfreeint}, one of our
 methods to access the surface free energy is to integrate
 the surface energy over $\beta$. Here we now describe
 how we did this in practice.

 We did simulations with antiperiodic boundary conditions
 in $z$-direction on lattices with $L=8,16,32,64$ for
 $\beta$-values ranging from the bulk critical region
 up to $\beta = 0.6$ which is deep in the low temperature
 domain. For many $\beta$-values we made runs with different
 $D$ to control the effects of a finite thickness of the
 lattice in $z$-direction. In total, we made more than
 250 different simulations with antiperiodic boundary
 conditions. Typically, we made 10000 measurements of
 several quantities, separated always by
 8 cluster updates (alternating of type O and B) and a single
 Metropolis sweep.
 A major part of the simulations was done on
 RISC workstations.

 A remark concerning the performance of the algorithms:
 The algorithm version C performed quite well for $\beta \geq 0.25$.
 The typical autocorrelation times (in units of the sweeps defined
 above) were of order one throughout. The type S worked well for
 all $\beta$, as long as the ratio of $T$ and the surface width
 was sufficiently small.

 The simulations supplied us with
 a sufficiently dense grid of $\beta$-values for the energies
 $E_{a.p.}$.

 For most of the $\beta$-values, we fortunately
 did not have to do extra simulations to access the
 energies with periodic boundary $E_{p.}$. Instead we
 used the diagonal
 Pad\'e approximation (order of nominator =  order of denominator
 = 12) of the low temperature
 series by Bhanot et al.\ \cite{epslow}, cf.\
 the appendix. By comparing with Monte Carlo
 simulations we found that this approximation is
 safe for a seizable range of $\beta$-values, cf.\
 table 1. In the table we quote
 the $\beta$-values above which we used the
 Pad\'e approximant throughout. For smaller $\beta$ values
 we used the Cluster Monte Carlo method to determine $E_{p.}$.
 To give an impression of the data we display
 our results for the surface energies (divided by
 $L^2$) in figure 1.

 In order to do the integration over $\beta$
 we interpolated the data with the help of
 cubic splines which can easily be integrated
 over arbritrary intervals numerically.
 Estimates for the statistical error of the surface free energy were
 obtained as follows. For each of the $\beta$-values,
 we have an energy value and an error bar. Note that
 the data for different $\beta$ are statistically
 independent. We simulated a whole sequence
 of outcomes ``energy as function of $\beta$'' by
 generating independent Gaussian random numbers centered around
 the Monte Carlo averages and with variances determined
 by the error bars, respectively.
 For each of these outcomes, a spline was generated and integrated.
 The error of the result of the integration (the free energy)
 was then obtained as the mean square deviation over this
 ``data Monte Carlo''.

 We employed this method to do the integration over $\beta$
 starting from large $\beta$ as well as from small $\beta$.
 The integration from large $\beta$ was always started
 at $\beta=0.6$, where the integration constant can be
 safely taken from the low temperature series for
 the surface tension. The ``initial conditions'' for
 the integration starting at small $\beta$ were
 already described in section \ref{SUBSUBSECfreeint}.

 The results for the free energies for $L=8,16,32$ and $64$
 were then used to make fits with the ansatz eq.\ (\ref{sigmaL})
 to obtain estimates for the surface tension $\sigma$.
 Two typical fits are displayed in figure 2.
 We determined $\sigma$'s both from the two different
 integration directions. A subset of our results is displayed
 in table 2. The $\chi^2$'s quoted in the last column show
 that the fits for the data from the integration started at
 small $\beta$ have a significantly higher $\chi^2$.
 We also made fits with the $L=8$ data excluded, and they
 had a better $\chi^2$/d.o.f. We conclude from this that close
 to the critical point the inclusion of larger lattices
 might be necessary to give reliable estimates for $\sigma$.
 Our method would allow us to do this.

 With figure 3, we demonstrate how nicely
 the results from the two integration directions match
 within the error bars. The constants $C_s$, however,
 show deviations indicating systematic effects which
 we do not have under control. This, of course, carries
 over to the approximate determination of the constant
 $C$ defined in equation (\ref{appro}).

 We also tried to estimate the critical exponent $\mu$ that
 governs the critical behavior of the surface tension.
 To this end we fitted our results for $\sigma$ according
 to the critical law $\sigma = \sigma_0 t^{\mu}$.
 Since our estimates for the $\sigma$'s at
 different $\beta$-values are strongly correlated
 we took the covariance matrix into account when
 doing the ``data Monte Carlo'' for the error estimates.
 We made two sorts of fits: Fits with the definition
 $t=1-\beta_c/\beta$ and fits with the definition
 $t=\beta/\beta_c - 1$. We also varied the interval, over
 which the the $\beta$ dependence of $\sigma$ was fitted.
 Our results are shown in tables 3,4 and 5.
 The fits were always done using four different $\beta$-values.
 Using more data points would not make very much sense since
 the data are correlated anyhow. Note however, that our
 statistical errors are nevertheless correct since our
 ``data Monte Carlo'' takes the covariances correctly into account.
 The comparison of the two different fits (using the two different
 definitions of $t$) clearly shows that there are systematic
 effects larger than the error bars: one still is not close enough
 to criticality. However we think that it is fair to say that
 our results are consistent with the value of $\mu \approx 1.26$ expected
 from Widom scaling. The results for the critical amplitude
 $\sigma_0$ show even stronger dependency on the type of the fit,
 and we can not say very much more that $\ln \sigma_0$ is probably
 something between $0.2$ and $0.4$.

 The method to compute surface free energies by
 finite $\Delta \beta$-steps worked quite well.
 In tables 6 and 7
 we present some of our results for the
 surface free energy on lattices with $L=8,16$,$32$ and $64$.
 We there display the naive free energy
 $F_{s,\rm n}(L)$ obtained by assuming only
 a single interface, and the ``improved'' free energy
 $F_{s,\rm i}(L)$ that is computed taking
 into account the presence of several interfaces.
 For small interface area and for $\beta$ close to the
 critical point the difference between the two definitions
 becomes significant.
 We also determined estimates for the surface tension
 $\sigma$ from the surface free energies.
 The results are quoted in table 8. Again the fits have
 relatively large $\chi^2$, and in some cases
 discarding the $L=8$ data
 changes the results beyond the statistical error. We again
 consider this as a warning
 that too small $L$'s might lead to systematic errors
 in $\sigma$.

\subsubsection{The correlation length in the broken phase}

 In order to study the behavior of
 the products $\xi^2 \sigma$ we tried also to
 extract the correlation length from the simulations of the systems
 with periodic boundary conditions.
 We define the magnetization of a time slice as
 \be
  S_{i_z} = \frac{1}{L^2} \sum_{i_x,i_y} s_{i} \, .
 \ee
 We define the following
 connected correlation function $G(t)$ of the time slice
 magnetizations:
 \be\label{corrG}
  G(t) = \langle
 | S_{i_z} S_{i_z+t} | \rangle
 - \lim_{\tau \rightarrow \infty}
  \langle  | S_{i_z} S_{i_z +\tau} | \rangle \, .
 \ee
  For our practical purposes we used $\tau = D$ which is the largest
  separation on the periodic lattices.
  For large $t$ the correlation function $G(t)$ is assumed to behave
  like
 \be
  G(t) \propto \exp(-\frac{t}{\xi}) \, .
 \ee
  (As a consequence of our definition of $G(t)$,
  $\xi$ can not rigorously be considered as the inverse
  of a physical mass. For finite $L$ the masses are split due
  to tunneling. However, the signal of tunneling
  is suppressed by the insertion
  of absolute values in eq.\ (\ref{corrG}).)

  We extracted estimates for $\xi$ from two subsequent values $t$ and
  $t+1$. One gets
 \be
 \xi_{\rm eff}(t,t+1) = - \frac{1}{\ln G(t) - \ln G(t+1)} \, .
 \ee
  The statistical error increases rapidly with increasing $t$. On the
  other hand the estimate for $\xi$
  is spoiled by systematic errors
  if we choose a too small $t$.
  As a compromise we chose the $t$ for our final estimate selfconsistently
  such that
  $t > \xi > t+1$. The results are quoted in table 9.
  One should remark that for all our data the lattice size was at least
  12 times the correlation length.

  We also tried to extract the correlation length from low temperature
  series \cite{Arisue,TarkoFisher}.
  Our analysis is based on the quantity
 \be \label{laeq}
  \Lambda_2 = \frac{ \exp(-\xi^{-1})}
                   { (1 - \exp(-\xi^{-1}))^2 } \, ,
 \ee
  where $\xi$ is the correlation length that controls the exponential
  decay of the correlation function.
  From the series for the correlation length
  given in \cite{Arisue}
  we derive the
  Taylor series for $\Lambda_2$,
 \begin{eqnarray}
  \Lambda_2 &=& u^2 - u^3 + 10 u^4 - 14 u^5
             + 93 u^6 - 201 u^7  \nonumber \\
            &+& \frac{4731}{7} u^8
             - \frac{33759}{56} u^9
             + \frac{115875}{28} u^{10}  \nonumber \\
            &-& \frac{295251}{56} u^{11}
             + \frac{4847861}{140} u^{12}
             - \frac{15341397}{280} u^{13}
             + \mbox{$\cal{O}$}(u^{14}) \, ,
  \end{eqnarray}
%
%
  where $ u = \exp(-4 \beta)$.
  To get reliable numbers from this series one has to control  the
  singularities. We proceed as follows.
  We first multiply the series  by
  $(u - 0.41205)^{1.25} (u + 0.336)$,
  where the first term stems from the
  critical singularity governed by the critical exponent $\nu$ that
  we assumed to equal 1.25 here. Variation of the input for
  $\nu$ in the range 0.625...0.63 led to changes in $\xi$ smaller
  than 3 per cent for $\beta=0.222$.
  The second term stems
  from an unphysical singularity
  on the negative real line.
  Then we Taylor expand the result and perform a
  Pad\'e aproximation on it (type [7,6]).
  We then divide
  the result of the Pad\'e approximation by
  $(u-0.41205)^{5/4} (u + 0.336)$.
  Finally we determine $\xi$ by inverting eq.\ (\ref{laeq}).
  In figure 4 we plot the quantity $\xi t^{\nu}$, with $\nu = 0.625$
  as obtained from the measured $\xi$'s and of the series result.
  Within the statistical accuracy the results are nicely consistent.

\subsection{Surface stiffness}
\label{SUBSECstiffmc}

 In all simulations we measured the block spin
 correlation functions
 $ A_{i,l}^{(\rm Ising)} $ as defined in section
 \ref{SUBSECstiff} for $i=1,2$ and $l=2,4$. We define
 auxiliary quantities
\be\label{beffhilf}
 \beta_{\rm eff}^{i,l} =
  \frac{ A_{i,l}^{(\rm Ising)} }{ A_{i,l}^{(0)} } \, ,
\ee
 where the $A^{(0)}$'s are taken for $L=256$
 (cf.\ table 10), which is essentially
 the infinite block size limit. We thus get eight values
 that all (in the large $L$ limit) should converge towards the
 same $\beff$. In table 11 we show two examples for
 these eight values (for two different values of the lattice
 thickness $T$) at $\beta=0.24$.
 The values for the different $i,l$ are fairly
 consistent within the statistical accuracy.
 Closer to the critical point the
 estimates from the bubble free configurations are more
 stable. We therefore decided to use only the bubble free
 data for the determination of the $\beta$-dependence of
 $\beff$.
 In tables 12 and 13 we
 present our estimates for $\beff$.
 The values were determined
 by averaging over the two quantities $\beta_{\rm eff}^{i,l}$ with
 $i=1$ and $i=2$. The first two lines of table 12 shows that
 our results become unstable close to the critical point where
 $\beff$ diverges.
 For $\beta = 0.43$ and $\beta = 0.45$, both points are in the
 smooth phase, $\beff$ decreases with increasing block size,
 in agreement with the Kosterlitz-Thouless picture of the
 roughening transition. This behavior is consistent with
 an infinite macroscopic stiffness for $\beta > \beta_r$.

 In figure 5, we show our results for two combined quantities,
 namely $\xi^2 \sigma$ and $\xi^2 \kappa$.
 In the product $\xi^2 \sigma$, the exponents $\mu$ and $-2\nu$
 of the reduced temperature $t$ should cancel, and we expect
 that this product should be fairly constant in a neighborhood of the
 critical point. The full line in figure 5 was obtained by combining
 our $\sigma's$ from the integration method with the correlation lengths
 as obtained from the Pad\'e.
 Since we do not know the error of the Pad\'e approximation of the
 low temperature series we base our error estimate
 for this quantity
 on our error bars for the measured correlation length as reported
 in table 9 and on the statistical errors on the surface tension $\sigma$.
 We estimate the relative precision of our results
 for $\xi^2 \sigma$ to be around 5 per cent for the smaller
 $\beta$'s, certainly better in the large $\beta$ region.
 This takes into account statistical errors only. There might
 also be systematic errors (due to too small $L$'s)
 in the surface tension close to
 the critical point. They might be responsible for another
 5 percent relative uncertainty.
 The points with error bars in figure 5 show the product
 $\xi^2 \kappa$. The plot shows that in the critical limit
 the surface stiffness becomes the same as the interface tension.
 This is a consequence of the restoration of rotational symmetry
 at the bulk critical point.
 In the theory of critical wetting, the following quantity plays
 an important role \cite{FisherWen}:
\be
 \omega(\beta) = 1 / (4 \pi \kappa \xi^2) \, .
\ee
 In \cite{FisherWen} we find an estimate $\omega(T/T_c = 0.8)=0.88$.
 Our result  $\omega(T/T_c = 0.8)=0.882(5)$ is in nice agreement
 with this prediction.
 In the limit $\beta \rightarrow \beta_r$, Kosterlitz-Thouless theory
 states that $\beff \rightarrow 2/\pi$. We use the estimate
 for $\beta_r$ cited in the introduction, and find
 $\xi(\beff) = 0.3163$ (from the Pad\'e that here certainly is reliable).
 We then find a ``KT-value'' of $\xi^2 \kappa$ which is 0.1572.


\subsection{Surface width}

 In order to demonstrate the efficiency of the Hasenbusch-Meyer
 algorithm we redid the surface width computation of Mon et al.\
 \cite{mon90a} at $ \beta = \beta_c/0.8 = 0.2771 $ on lattices of
 size $L \times L \times 27$,
 with $L=32,64,\dots,512$. Our results for
 $W^2$ and $W_0^2$ are summarized in table 14.

 We performed fits of the data using the ansatz
\be
W^2 = \mbox{const} + \frac{\beff}{2\pi} \ln(L)
\ee
 that is motivated by Kosterlitz-Thouless theory of a rough interface
 and that should become precise for large enough $L$ (depending, of course,
 on $\beta$). We performed several fits on subsets of the data.
 Our results for $\beff$ are summarized in table 15.
 The errors were determined by a ``data Monte Carlo'' as described
 in section \ref{SUBSECmcfe}.
 We conclude that
 the $L=32$ data should not be included in the fit and estimate
 that $\beff = 4.3(2)$.

 The result has to be compared with the $\beff$ as obtained from
 the renormalization group quantities $A$ introduced in
 section \ref{SUBSECstiff}. Because of the moderate statistics
 we used only quantities measured on the bubble free configurations.
 We computed the auxiliary quantities
 $\beta_{\rm eff}^{i,l} $, as defined in eq.\ (\ref{beffhilf}),
 measured on bubble free configurations only.
 The quantities $ \beta_{\rm eff}^{i,l} $ should converge
 to $\beff$ for large $L$.
 Our findings are summarized in table 16.
 One should not overemphasize the $L=512$-results,
 which suffer a bit from poor statistics. Instead we focus
 on the $L=128$ and on the $L=256$ results.
 Within the statistical accuracy these data are fairly
 compatible with the value for $\beff$ obtained from
 the surface thickness fits. We interpret our results as a further
 confirmation that the long distance properties of the Ising interface
 are correctly described by a massless Gaussian dynamics.

\section{Conclusions}\label{SECconclu}

 In this paper, we have presented a numerical study of the Ising interface
 in three dimensions over a wide range of temperatures. The method
 to obtain the surface free energies by integration over $\beta$
 requires many separate simulations but turns out to be practicable
 and useful. Inclusion of large interfaces is
 possible because of the usage of a cluster algorithm also for the
 simulations with antiperiodic boundary conditions.

 Our analysis of the surface tension indeed showed that
 closer to the critical point large interface extensions $L$ are
 necessary to get reliable values ($\sigma$ has a tendency
 to come out too large when the lattices are too small).
 A high precision computation of $\sigma_0$ and $\mu$ seems difficult,
 the systematic effects from a too large reduced temperature $t$
 are strong.

 The large interfaces allowed us to study the infrared
 surface properties. We could confirm
 the massless Gaussian behavior in the rough
 phase and also extract the stiffness coefficient from
 the renormalization group behavior.

\section*{Acknowledgements}
 One of us (M.H.) would like to acknowledge
 support by Deutsche Forschungsgemeinschaft
 through grant \#Me\ 567/5-3.
 A useful discussion with Gernot M\"unster is gratefully acknowledged.
 We would like to thank
 the computer center of the University in M\"unster and
 the RHRK (Regionales Hochschulrechenzentrum Kaiserslautern)
 where most of the computer simulations where done.

\newpage
\section*{Appendix: Low temperature series for energy and surface tension}

 For the convenience of the reader, we here cite the low temperature
 series for the energy density
 and for the surface tension.

 Bhanot et al.\ \cite{epslow} recently pushed forward
 the low temperature series for the energy
 to $24^{\mbox{\small th}}$ order in the variable
 $u = \exp(-4 \beta)$. One defines
\be
 \epsilon = 3 ( 1 -  \langle \sigma_i \sigma_j \rangle)
 \mbox{\ \ for $i,j$ nearest neighbors.}
\ee
 The series for $\epsilon$ is
\begin{eqnarray}
 \epsilon &=&
   12 \, u^3
 + 60 \, u^5
 - 84 \, u^6
 + 420 \, u^7
 - 1056 \, u^8
 \nonumber \\
 &+& 3756 \, u^9
 - 11220 \, u^{10}
 + 37356 \, u^{11}
 - 118164 \, u^{12}
 + 389220 \, u^{13}
 \nonumber \\
 &-& 1261932 \, u^{14}
 + 4163592 \, u^{15}
 - 13680288 \, u^{16}
 + 45339000 \, u^{17}
 \nonumber \\
 &-& 150244860 \, u^{18}
 + 500333916 \, u^{19}
 - 1668189060 \, u^{20}
 \nonumber \\
 &+& 5579763432 \, u^{21}
 - 18692075820 \, u^{22}
 \nonumber \\
 &+& 62762602860 \, u^{23}
 - 211062133044 \, u^{24}
 + \mbox{$\cal{O}$} (u^{25}) \, .
\end{eqnarray}

A low temperature series to order $u^9$
for the surface tension $\sigma$ was determined
by Weeks et al.\ The coefficients can be found in a paper by
Shaw and Fisher \cite{ShawFisher}. The series is

\begin{eqnarray}
 \sigma &=& 2 \, \beta
  - 2 \, u^2
  - 2 \, u^3
  - 10 \, u^4
  - 16 \, u^5
 \nonumber \\
  &-& \frac{242}{3} \, u^6
  - 150 \, u^7
  - 734 \, u^8
  - \frac{4334}{3} \, u^9
  + \mbox{$\cal{O}$} (u^{10}) \, .
\end{eqnarray}



\newpage
\listoftables

\newpage
\section*{List of Figures}

\noindent
{\bf Figure 1:} The results for the surface energy per area as
function of $\beta$ for spatial lattice extension
$L=8,16,32,64$. These are the data to be interpolated by splines
and integrated over in order to determine the surface free energies.

\vskip1cm
\noindent
{\bf Figure 2:} Two examples for our results for the surface free energy
as function of $L$. These data are fitted with the law
$F_s = C_s + \sigma L^2$ to determine the surface tension.

\vskip1cm
\noindent
{\bf Figure 3:} With this figure, we demonstrate the consistency of
the results for $\sigma$ obtained
from the integration started at large $\beta$ with that obtained from
the integration started at small $\beta$. The left half of the
figure contains only the ``small $\beta$'' data, whereas the right
part contains only the ``large $\beta$'' data. Here they meet
at $\beta = 0.235$. (The data from both methods
cover the whole $\beta$ range. We chose this specific presentation
only to demonstrate the consistency.) For small $\beta$, the $\sigma's$
obtained by integrating from below have smaller error bars.

\vskip1cm
\noindent
{\bf Figure 4:} Comparison of our results for the correlation length
in the broken phase with the Pad\'e approximation of the low temperature
series of ref.\ \cite{Arisue}. The figure shows the result for the
quantity $\xi t^{\nu}$, with $\nu = 0.625$.

\vskip1cm
\noindent
{\bf Figure 5:} Plot of our results for the
combined quantities $\xi^2 \sigma$ (full line) and
$\xi^2 \kappa$ (points with error bars). Details are
explained in the text.

\newpage

\begin{table}\label{compMCPADE}
\caption{Comparison of Monte Carlo results for
         energy per site ($=3\langle \sigma_i \sigma_j \rangle$, with
         $i$ and $j$ nearest neighbors, periodic boundary conditions)
         with results from the Pad\'e approximation of the low temperature
         series. In the last column we quote the $\beta$ value
         above which we consider the usage of the Pad\'e approximation
         as safe.}
\begin{center}
\begin{tabular}{|c|c|c|l|c|c|}
\hline
   $L$ &  $D$  &  $\beta$  &
   \multicolumn{1}{c|}{MC}   &     Pad\'e   &  Pad\'e used for ... \\
\hline
   8   &   9   &  0.2350   & 1.5381(8)  &  1.5472 &  \\
   8   &   9   &  0.2400   & 1.6826(7)  &  1.6844 &  \\
   8   &   8   &  0.2500   & 1.9082(20) &  1.9083 &  $\beta \geq 0.26$ \\
\hline
  16   &  16   &  0.2300   & 1.3849(6)  &  1.3860 &  \\
  16   &  16   &  0.2325   & 1.4702(7)  &  1.4701 &  \\
  16   &  16   &  0.2350   & 1.5469(6)  &  1.5472 &  $\beta \geq 0.24$ \\
\hline
  32   &  16   &  0.2300   & 1.3839(9)  &  1.3860 &  \\
  32   &  16   &  0.2327   & 1.4764(8)  &  1.4765 &  \\
  32   &  16   &  0.2350   & 1.5466(8)  &  1.5472 &  $\beta \geq 0.24$ \\
\hline
  64   &  16   &  0.2265   &  1.2498(6) &  1.2537 &  \\
  64   &  16   &  0.2275   &  1.2901(6) &  1.2935 &  \\
  64   &  16   &  0.2300   &  1.3852(5) &  1.3860 &  $\beta \geq 0.235$ \\
\hline
\end{tabular}
\end{center}
\end{table}

\begin{table}\label{subsigma}
\caption{Results for the constant $C_s$ and for the
         surface tension $\sigma$. Type l means: obtained
         by integration starting at large $\beta$, type s means: obtained by
         integration starting at small $\beta$.
         We also quote the logarithm of the constant $C$ defined in
         section 3.2.3. The last column gives
         $\chi^2$ per degree of freedom, averaged over the
         ``data Monte Carlo''.}
 \begin{center}
 \begin{tabular}{|c|c|c|l|c|c|}
\hline
$\beta$ & type & $C_s$ &
\multicolumn{1}{c|}{$\sigma$} & $ - \ln C$  & $\chi^2$/d.o.f \\
\hline
                            & l & 2.885(81) & 0.00733(12) & 2.192(81) &  1.09
\\
\rbo{1.5ex}[-1.5ex]{0.225}  & s & 2.663(20) & 0.007489(38) & 1.970(20)&  7.45
\\
\hline
                           & l & 2.770(83) &  0.01014(13) &2.077(83) &  1.33
\\
\rbo{1.5ex}[-1.5ex]{0.226} & s& 2.539(22) &   0.010318(44) &1.846(22) &  5.35
\\
\hline
                           & l & 2.651(77) &  0.01310(12)  &1.958(77)& 1.27 \\
\rbo{1.5ex}[-1.5ex]{0.227} & s & 2.425(21) &   0.013303(52) &1.732(21)& 4.65 \\
\hline
 & l & 2.554(76) &  0.01620(12) &1.861(76)&  1.39 \\
\rbo{1.5ex}[-1.5ex]{0.228} & s & 2.324(21)  &  0.016406(55) &1.631(21)&  6.16
\\
\hline
 & l& 2.477(81) &  0.01938(11) &1.784(81)&  1.62 \\
\rbo{1.5ex}[-1.5ex]{0.229} & s& 2.235(22) &   0.019589(56) &1.542(22)&  7.43 \\
\hline
 & l  & 2.388(73) &  0.02268(11) &1.695(73)&  1.84 \\
\rbo{1.5ex}[-1.5ex]{0.230} & s& 2.151(21)  &  0.022863(68) &1.458(21)&  6.46 \\
\hline
 & l &2.298(80) &  0.02606(11) &1.605(80)&  1.72 \\
\rbo{1.5ex}[-1.5ex]{0.231} & s& 2.073(24) &   0.026226(63) &1.380(24)&  5.82 \\
\hline
 &  l& 2.232(76) &  0.02952(11)&1.539(76)&   1.27 \\
\rbo{1.5ex}[-1.5ex]{0.232} &  s& 1.999(26) &   0.029684(82) &1.306(26)&  5.03
\\
\hline
 & l& 2.165(77)  & 0.03305(11) &1.472(77)&  1.25 \\
\rbo{1.5ex}[-1.5ex]{0.233} &  s& 1.934(25)  &  0.033221(82) &1.241(25)&  5.27
\\
\hline
 & l& 2.097(73) &  0.03665(11) &1.404(73)&  1.20 \\
\rbo{1.5ex}[-1.5ex]{0.234} & s& 1.877(26) &   0.036808(89) &1.184(26)&  5.24 \\
\hline
 & l & 2.048(69) &  0.04025(10) &1.355(69)&  1.14 \\
\rbo{1.5ex}[-1.5ex]{0.235} & s& 1.831(24) &   0.040429(84) &1.138(24)&  5.25 \\
\hline
 & l& 1.999(72) &  0.04392(11)&1.306(72)&   1.13 \\
\rbo{1.5ex}[-1.5ex]{0.236} &  s& 1.785(31)  &  0.044077(95) &1.092(31)& 5.31
\\
\hline
 & l& 1.947(71)&   0.04760(10) &1.281(71)&  1.01 \\
\rbo{1.5ex}[-1.5ex]{0.237} &  s& 1.747(32) &   0.047758(85) &1.054(32) & 5.68
\\
\hline
 & l& 1.909(75) &  0.05133(10) &1.216(75)&  1.15 \\
\rbo{1.5ex}[-1.5ex]{0.238} & s & 1.707(35) &   0.05149(10) &1.014(35)&  5.69 \\
\hline
 & l & 1.869(70) &  0.05509(10) &1.176(70)&  1.08 \\
\rbo{1.5ex}[-1.5ex]{0.239} & s  &1.671(30)  &  0.05524(10)  &0.978(30)&  5.35
\\
\hline
 & l & 1.840(66) &  0.05889(10) &1.147(66)&  1.11 \\
\rbo{1.5ex}[-1.5ex]{0.240} & s & 1.636(33)  &  0.05905(11) &0.943(33)& 5.68 \\
\hline
   \end{tabular}
  \end{center}
 \end{table}

\begin{table}\label{mutable1}
\caption{Results for the critical exponent $\mu$ as obtained from
         fitting the logarithm of the surface tension with the critical
         law $\ln \sigma = \ln \sigma_0 + \mu \ln t$.
         For this table we used the
         $\sigma$'s from the integration starting at {\em large} $\beta$.
         Type 1 is a fit with $ t = 1-\beta_c / \beta$, and
         type 2 is a fit with $t= \beta / \beta_c - 1 $.
         The fourth and fifth columns give the results using
         $\sigma$'s obtained only from subsets of the data
         ($L=8$ data excluded or $L=64$ data excluded, respectively).
         The fits were done using four equidistant $\beta$-values.
         The numbers in square brackets give the average
         $\chi^2$ in the ``data Monte Carlo'', not divided by d.o.f.}
\begin{center}
\begin{tabular}{|c|c|l|c|c|}
\hline
type of fit & fit interval &
\multicolumn{1}{c|}{all lattice sizes} & $ -(L=8) $ & $-(L=64)$ \\
 \hline
1 &  & 1.274(16) [0.27] & 1.276(19) [0.52] & 1.266(31) [0.13] \\
2 & \rbo{1.5ex}[-1.5ex]{0.224 - 0.2300}
  & 1.243(15) [0.57] & 1.244(18) [0.90] & 1.234(30) [0.08] \\
\hline
1 &  & 1.270(12) [0.21] & 1.275(17) [0.35] & 1.257(28) [0.19] \\
2 & \rbo{1.5ex}[-1.5ex]{0.224 - 0.2313}
  & 1.234(11) [0.48] & 1.239(17) [0.72] & 1.222(27) [0.20] \\
\hline
1 &  & 1.278(9) [0.38]  & 1.282(12) [0.67] & 1.252(22) [0.25] \\
2 & \rbo{1.5ex}[-1.5ex]{0.224 - 0.2325}
  & 1.238(9) [1.18]  & 1.241(11) [1.38] & 1.213(21) [0.33] \\
\hline
1 &  & 1.277(8) [0.25]  & 1.281(11) [0.52] & 1.259(21) [0.28] \\
2 & \rbo{1.5ex}[-1.5ex]{0.224 - 0.2338}
  & 1.232(7) [1.02]  & 1.236(19) [1.33] & 1.216(20) [0.34] \\
\hline
1 &  & 1.276(8) [0.57]  & 1.281(11) [0.59] & 1.248(18) [0.43] \\
2 & \rbo{1.5ex}[-1.5ex]{0.224 - 0.2350}
  & 1.226(8) [0.96]  & 1.232(19) [1.24] & 1.200(17) [0.25] \\
\hline
\end{tabular}
\end{center}
\end{table}

\begin{table}\label{mutable2}
\caption{Same as table 3, however based on results for $\sigma$ obtained
         from the integration starting at {\em small} $\beta$.}
\begin{center}
\begin{tabular}{|c|c|c|l|c|}
\hline
type of fit & fit interval & all lattice sizes &
\multicolumn{1}{c|}{$ -(L=8)$} & $-(L=64)$ \\
 \hline
1 &  & 1.252(5) [0.24] & 1.268(7) [3.63] & 1.240(7) [3.27] \\
2 & \rbo{1.5ex}[-1.5ex]{0.224 - 0.2300}
  & 1.223(5) [1.07] & 1.238(7) [7.41] & 1.212(7) [1.21] \\
\hline
1 &  & 1.253(4) [0.56] & 1.266(5) [3.3]  & 1.242(7) [3.40] \\
2 & \rbo{1.5ex}[-1.5ex]{0.224 - 0.2313}
  & 1.221(4) [1.30] & 1.233(5) [8.5]  & 1.211(7) [1.31] \\
\hline
1 &  & 1.254(4) [2.18] & 1.268(5) [2.79] & 1.244(7) [4.57] \\
2 & \rbo{1.5ex}[-1.5ex]{0.224 - 0.2325}
  & 1.219(4) [0.87] & 1.231(4) [8.58] & 1.210(6) [1.72] \\
\hline
1 &  & 1.257(4) [5.01] & 1.270(5) [2.96] & 1.246(6) [5.87] \\
2 & \rbo{1.5ex}[-1.5ex]{0.224 - 0.2338}
  & 1.218(4) [0.86] & 1.230(5) [9.34] & 1.208(5) [1.40] \\
\hline
1 &  & 1.261(3) [3.90] & 1.269(3) [2.48] & 1.249(5) [4.56] \\
2 & \rbo{1.5ex}[-1.5ex]{0.224 - 0.2350}
  & 1.218(3) [1.83] & 1.226(3) [11.11]& 1.208(5) [0.70] \\
\hline
\end{tabular}
\end{center}
\end{table}

\begin{table}\label{contable1}
\caption{Results for $\ln \sigma_0$ as obtained from
         fitting the logarithm of
         the surface tension $\ln \sigma$ with the critical
         law $\ln \sigma = \ln \sigma_0 + \mu \ln t$.
         For this table we used the
         $\sigma$'s from the integration starting at {\em small} $\beta$.
         Type 1 is a fit with $ t = 1-\beta_c / \beta$, and
         type 2 is a fit with $t= \beta / \beta_c - 1 $.
         The fourth and fifth columns give the results using
         $\sigma$'s obtained only from subsets of the data
         ($L=8$ data excluded or $L=64$ data excluded, respectively).
         The fits were done using four equidistant $\beta$-values.
         The numbers in square brackets give the average
         $\chi^2$ in the ``data Monte Carlo'', not divided by d.o.f.}
\begin{center}
\begin{tabular}{|c|c|c|l|c|}
\hline
type of fit & fit interval & all lattice sizes &
\multicolumn{1}{c|}{$ -(L=8)$} & $-(L=64)$ \\
 \hline
1 &  & 0.372(18) [0.24] & 0.423(22) [3.63] & 0.341(23) [3.27] \\
2 & \rbo{1.5ex}[-1.5ex]{0.224 - 0.2300}
  & 0.232(17) [1.07] & 0.281(22) [7.41] & 0.206(22) [1.21] \\
\hline
1 &  & 0.376(15) [0.56] & 0.416(17) [3.30] & 0.349(25) [3.39] \\
2 & \rbo{1.5ex}[-1.5ex]{0.224 - 0.2313}
  & 0.223(15) [1.30] & 0.261(17) [8.52] & 0.239(24) [1.31] \\
\hline
1 &  & 0.383(12) [2.17] & 0.422(14) [2.79] & 0.357(23) [4.57] \\
2 & \rbo{1.5ex}[-1.5ex]{0.224 - 0.2325}
  & 0.216(11) [0.87] & 0.253(13) [8.58] & 0.198(21) [1.72] \\
\hline
1 &  & 0.392(11) [5.01] & 0.429(15) [2.96] & 0.364(18) [5.87] \\
2 & \rbo{1.5ex}[-1.5ex]{0.224 - 0.2338}
  & 0.214(11) [0.85] & 0.248(14) [9.34] & 0.192(16) [1.39] \\
\hline
1 &  & 0.404(10) [3.90] & 0.427(9) [2.48]  & 0.374(17) [4.56] \\
2 & \rbo{1.5ex}[-1.5ex]{0.224 - 0.2350}
  & 0.213(10) [1.83] & 0.234(9) [11.11] & 0.189(16) [0.70] \\
\hline
\end{tabular}
\end{center}
\end{table}

\begin{table}\label{fevo1}
\caption{Surface free energies obtained by the step-by-step
         method, for $L=8$ and $L=16$. We here display the naive free energy
         (subscript $s$)
         obtained by assuming only
         a single interface, and the ``improved'' free energy
         (subscript $i$) that is computed taking
         into account the presence of several interfaces.}
\begin{center}
\begin{tabular}{|c||c|c||c|c|}
\hline
$\beta$ & $F_{s,\rm n}(8)  $
        & $F_{s,\rm i}(8)  $
        & $F_{s,\rm n}(16) $
        & $F_{s,\rm i}(16) $ \\
\hline
0.222 & 3.026(12)  &  2.475(43)
       & 3.688(19)  & 3.337(40)   \\
0.223 & 3.060(14)  &  2.589(41)
       & 3.882(24)  & 3.684(36)   \\
0.224 & 3.103(16)  &  2.706(39)
       & 4.200(30)  & 4.108(36)   \\
0.225 & 3.156(19)  &  2.827(39)
       & 4.649(32)  & 4.614(40)   \\
0.226 & 3.228(19)  &  2.964(33)
       & 5.208(35)  & 5.197(36)   \\
0.227 & 3.311(20)  &  3.103(30)
       & 5.849(37)  & 5.846(37)   \\
\hline
\end{tabular}
\end{center}
\end{table}

\begin{table}\label{fevo2}
\caption{Same as table 6, however for $L=32$ and $L=64$.}
\begin{center}
\begin{tabular}{|c||r|r||r|r|}
\hline
$\beta$ & \multicolumn{1}{|c|}{$F_{s,\rm n}(32) $}
        & \multicolumn{1}{c|}{$F_{s,\rm i}(32) $}
        & \multicolumn{1}{|c|}{$F_{s,\rm n}(64) $}
        & \multicolumn{1}{c|}{$F_{s,\rm i}(64) $} \\
\hline
0.222 & 4.340(28)  &  4.272(32) &    5.82(11) &   5.82(11) \\
0.223 & 5.706(36)  &  5.702(37) &   12.19(15) &  12.19(15) \\
0.224 & 7.841(43)  &  7.841(43) &   21.90(17) &  21.90(17) \\
0.225 & 10.386(48) & 10.386(48) &   32.78(19) &  32.78(19) \\
0.226 & 13.179(52) & 13.179(52) &   44.21(20) &  44.21(20) \\
0.227 & 16.123(56) & 16.123(56) &             &            \\
\hline
\end{tabular}
\end{center}
\end{table}
 \begin{table}\label{sigextra}
 \caption{Results of the fit of the form $F_{im} = C_s + \sigma L^2$
   for the data obtained by the step-by-step method. The last column
   gives $\chi^2$ per degrees of freedom.}
   \begin{center}
   \begin{tabular}{|c|c|c|c|c|}
  \hline
    $\beta$ &L's used& $C_s$ & $\sigma$ & $\chi^2/\mbox{d.o.f.}$ \\
   \hline
      0.227  & 8,16,32 &2.28(3) & 0.01356(7) &10.7 \\
   \hline
      0.227  &   16,32 &2.42(5) & 0.01338(9) &     \\
   \hline
      0.226   &16,32,64 &2.61(4) &0.01023(5)& 6.2     \\
   \hline
      0.225   &16,32,64 &2.76(4) &0.00738(4)& 4.0     \\
   \hline
      0.224  & 16,32,64 & 2.96(4) & 0.00469(4) & 8.   \\
   \hline
      0.224  &    32,64 & 3.16(8) & 0.00457(6) &      \\
   \hline
      0.223 &16,32,64 & 3.20(4) & 0.00231(4) & 34. \\
   \hline
      0.223 &   32,64 & 3.54(7) & 0.00211(5) &     \\
   \hline
   \end{tabular}
  \end{center}
 \end{table}

\begin{table}\label{xitable}
\caption{Results for the correlation length $\xi$. $t$ denotes the
 distance at which the correlation length is determined.}
 \begin{center}
 \begin{tabular}{|l|c|c|c|l|}
\hline
 \multicolumn{1}{|c|}{$\beta$} & $L$ & $T$  & $t$ &
 \multicolumn{1}{c|}{$\xi$} \\
 \hline
   0.2250   & 64 & 65 & 3-4 &  3.57(10) \\
   0.2275   & 64 & 33 & 2-3 &  2.45(7)  \\
   0.2285   & 32 & 33 & 2-3 &  2.16(6)  \\
   0.2300   & 64 & 33 & 2-3 &  2.04(6)  \\
   0.2327   & 32 & 33 & 1-2 &  1.66(3)  \\
   0.2350   & 32 & 33 & 1-2 &  1.48(3)  \\
   0.2391   & 16 & 23 & 1-2 &  1.28(3)  \\
   0.2400   & 16 & 23 & 1-2 &  1.22(3)  \\
   0.2500   & 16 & 23 & 0-1 &  0.92(1)  \\
   0.2600   & 16 & 11 & 0-1 &  0.76(1)  \\
   0.2700   & 16 & 11 & 0-1 &  0.68(1)  \\
   0.2800   & 16 & 11 & 0-1 &  0.61(1)  \\
   0.2900   & 16 & 11 & 0-1 &  0.55(1)  \\
   0.3000   & 16 & 11 & 0-1 &  0.52(1)  \\
   0.3200   &  8 & 11 & 0-1 &  0.46(1)  \\
   0.3400   &  8 & 11 & 0-1 &  0.40(1)  \\
   0.3600   &  8 & 11 & 0-1 &  0.36(1)  \\
 \hline
   \end{tabular}
  \end{center}
 \end{table}

\begin{table}\label{freeblock}
\caption{Exact results for $A_{i,l}^{(0)}$ for $\beta=1$ as obtained
         by Fourier transformation.
         $L$ denotes the size of the 2-dimensional lattice.}
   \begin{center}
   \begin{tabular}{|r|c|c|c|c|c|}
   \hline
 \multicolumn{1}{|c|}{ $L$ }
 & $A_{1,2}^{(0)}$ & $A_{2,2}^{(0)}$
       & $A_{1,4}^{(0)}$ & $A_{2,4}^{(0)}$     \\
   \hline
   4 & 0.187500 & 0.250000 &          &          \\
   8 & 0.136719 & 0.187500 & 0.293527 & 0.380581 \\
  12 & 0.126721 & 0.175926 & 0.257225 & 0.340481 \\
  16 & 0.123147 & 0.171875 & 0.243918 & 0.326090 \\
  24 & 0.120565 & 0.168981 & 0.234147 & 0.315663 \\
  32 & 0.119655 & 0.167969 & 0.230662 & 0.311978 \\
  48 & 0.119002 & 0.167245 & 0.228148 & 0.309333 \\
  64 & 0.118773 & 0.166992 & 0.227263 & 0.308404 \\
  96 & 0.118609 & 0.166811 & 0.226628 & 0.307739 \\
 128 & 0.118551 & 0.166748 & 0.226406 & 0.307507 \\
 256 & 0.118496 & 0.166687 & 0.226191 & 0.307282 \\
   \hline
   \end{tabular}
  \end{center}
 \end{table}
\begin{table}\label{beffdemo}
\caption{The auxiliary quantities $\beta_{\rm eff}^{i,l}$
         for $\beta=0.24$ on a $64 \times 64 \times T$ lattice,
         with (b) and without (nb) bubbles.}
\begin{center}
\begin{tabular}{|c|c|c|c|c|c|}
\hline
$i$ & $l$ & (b) $T=19$ & (nb) $T=19$
& (b) $T=27$ & (nb) $T=27$ \\
\hline
 1  &  2  & 16.88(44) & 16.98(47) &   16.94(47)  &  16.59(46)   \\
 2  &  2  & 16.82(52) & 16.85(52) &   16.27(56)  &  16.06(55)   \\
 1  &  4  & 16.82(20) & 17.18(20) &   17.64(22)  &  16.86(21)   \\
 2  &  4  & 16.79(23) & 17.04(23) &   17.35(25)  &  16.79(24)   \\
\hline
\end{tabular}
\end{center}
\end{table}
\begin{table}\label{beff1}
\caption{ $\beta_{\rm eff}$ for $\beta=0.24 \dots 0.32$ as obtained
 from $2 \times 2$ and $4 \times 4$ blocking on lattices with $L=16,32,64$.
 Only bubble free configuration were used. For each value of $l$, the
 estimate for $\beff$ was obtained by taking the average of
 $\beta_{\rm eff}^{1,l}$ and
 $\beta_{\rm eff}^{2,l}$.}
\begin{center}
\begin{tabular}{|c||c|c||c|c||c|c|}
\hline
           & \multicolumn{2}{c|}{ $L=16$}
           & \multicolumn{2}{c|}{ $L=32$}
           & \multicolumn{2}{c|}{ $L=64$} \\
\hline
   $\beta$ & $l=2 $ & $l=4 $ & $l=2 $ & $l=4 $ & $l=2 $ & $l=4 $  \\
\hline
0.230&           &         &  62.3(36) & 73.6(89) & 45.5 (20) & 45.59(86) \\
0.235& 28.84(61) & 28.30(43)& 28.67(42) &28.65(20) & 24.50(72)& 25.16(23) \\
\hline
0.240&  16.54(22) & 16.02(11)& 16.99(27) & 17.01(12) & 16.32(50) & 16.83(22) \\
0.245&          &         & 12.24(19) & 12.42(8) & 12.46(32) & 12.47(13) \\
0.250&  9.80(13) &  9.69(7) &  9.68(17) &  9.74(7) &  9.91(23) & 10.00(10) \\
0.255&          &         &  8.26(14) &  8.12(5) &  8.02(17) &  8.11(7) \\
0.260&  6.75(9) &  6.90(5) &  6.89(12) &  6.87(5) &  6.83(14) &  6.85(6) \\
0.265&          &         &  5.78(10) &  5.95(5) &  5.88(11) &  5.92(5) \\
0.270&  5.25(7) &  5.31(3) &  5.19(10) &  5.23(4) &  5.26(10) &  5.18(4) \\
0.275&          &         &  4.57(8) &  4.63(3) &  4.65(8) &  4.65(4) \\
0.280&          &         &  4.05(8) &  4.16(3) &  4.13(7) &  4.14(3) \\
0.285&  3.81(6) &  3.89(2) &  3.83(7) &  3.81(3) &  3.79(7) &  3.80(3) \\
0.290&          &         &  3.57(7) &  3.52(3) &  3.43(6) &  3.41(2) \\
0.295&          &         &  3.21(5) &  3.19(2) &  3.17(5) &  3.19(2) \\
0.300&  2.97(5) &  3.01(2) &  2.87(5) &  2.93(2) &  2.96(5) &  2.93(2) \\
0.305&          &         &  2.71(4) &  2.73(2) &  2.72(4) &  2.70(2) \\
0.310&          &         &  2.55(4) &  2.53(2) &  2.55(4) &  2.51(2) \\
0.315&          &         &  2.37(3) &  2.36(1) &  2.33(4) &  2.34(2) \\
0.320&          &         &  2.15(3) &  2.18(1) &  2.15(4) &  2.18(2) \\
 \hline
 \end{tabular}
 \end{center}
 \end{table}
\begin{table}\label{beff2}
\caption{ $\beta_{\rm eff}$ for $\beta=0.325 \dots 0.45$ as
 obtained from $2 \times 2$ and $4 \times 4$
 blocking on lattices with $L=16,32,64$.}
\begin{center}
\begin{tabular}{|c||c|c||c|c||c|c|}
\hline
           & \multicolumn{2}{c|}{ $L=16$}
           & \multicolumn{2}{c|}{ $L=32$}
           & \multicolumn{2}{c|}{ $L=64$} \\
\hline
   $\beta$ & $l=2 $ & $l=4 $ & $l=2 $ & $l=4 $ & $l=2 $ & $l=4 $  \\
\hline
0.325&  2.04(3) &  2.11(1) &  2.03(3) &  2.07(1) &  2.05(3) &  2.03(1) \\
0.330&          &         &  1.93(3) &  1.95(1) &  1.94(3) &  1.92(1) \\
0.335&          &         &  1.80(3) &  1.81(1) &  1.79(3) &  1.82(1) \\
0.340&          &         &  1.71(2) &  1.73(1) &  1.72(3) &  1.70(1) \\
0.345&          &         &  1.58(2) &  1.62(1) &  1.58(2) &  1.60(1) \\
0.350&  1.57(3) &  1.60(1) &  1.53(2) &  1.53(1) &  1.51(2) &  1.52(1) \\
0.355&          &         &          &         &  1.40(2) &  1.42(1) \\
0.360&          &         &          &         &  1.32(2) &  1.35(1) \\
0.365&          &         &          &         &  1.23(2) &  1.27(1) \\
0.370&  1.20(1) &  1.28(1) &  1.19(1) &  1.22(1) &  1.19(2) &  1.20(1) \\
0.375&          &         &          &         &  1.13(2) &  1.13(1) \\
0.385&          &         &          &         &  1.00(1) &  1.01(1) \\
0.390&  0.94(1) &  1.00(1) &  0.93(1) &  0.96(1) &  0.94(1) &  0.94(1) \\
0.405&          &         &          &         &  0.74(1) &  0.74(0) \\
0.410&  0.70(1) &  0.76(1) &  0.66(1) &  0.69(1) &          &         \\
0.430&  0.44(1) &  0.50(0) &  0.35(1) &  0.39(0) &  0.22(0) &  0.27(0) \\
0.450&  0.24(0) &  0.31(0) &  0.13(0) &  0.17(0) &  0.05(0) &  0.08(0) \\
 \hline
   \end{tabular}
  \end{center}
 \end{table}
\begin{table}\label{tabthick}
\caption{Squared surface width $W^2$ (measured on configurations with
         bubbles) and $W_0^2$ (measured after removal of the bubbles)
         at $\beta=\beta_c/0.8$ on lattices $L \times L \times 27$.
         $stat$ denotes the number
         of measurements. Between two subsequent measurements we always
         performed eight single cluster updates (alternating type O and B)
         and a single Metropolis sweep.
         For $L=256$ and $L=512$ we show results of two
         independent runs.}
\begin{center}
\begin{tabular}{|c|c|c|c|c|c|}
\hline
                & $L=32$  &  $L=64$  &  $L=128$   &  $L=256$  & $L=512$  \\
\hline
$W^2$           & 2.341(42) & 2.889(42) & 3.365(29) & 3.855(35) & 4.313(60) \\
                &           &           &           & 3.811(33) & 4.335(64) \\
\hline
$W_0^2$         & 3.094(9) & 3.619(16) & 4.098(20) & 4.573(35) & 5.059(66)  \\
                &          &          &            & 4.542(30) & 5.069(58)  \\
\hline
$stat$          & 10000    & 3000     & 2430       & 550 / 820 &  187/260  \\
\hline
\end{tabular}
\end{center}
\end{table}
\begin{table}\label{thickfit}
\caption{Fit results for $\beta_{\rm eff}$. The numbers in
         square brackets give
         the average $\chi^2$ per degree of freedom in the
         ``data Monte Carlo''.}
\begin{center}
\begin{tabular}{|c||l|l|l|l|}
\hline
data used           &  all  & all -- ($L=32$) & all -- ($L=32,64$) &
                       all -- ($L=512$) \\
\hline
\hline
from $W^2$          & 4.45(12) [1.33] & 4.30(16) [1.05]  &
                      4.30(23) [1.22] & 4.44(14) [1.87]  \\
\hline
from $W_0^2$        & 4.48(6)  [2.29]  & 4.3(1)    [1.31]  &
                      4.3(2)   [1.70]  & 4.48(6)   [3.62]   \\
\hline
\end{tabular}
\end{center}
\end{table}
%
\begin{table}\label{beffstauff}
\caption{$\beta_{\rm eff}^{i,l}$ for $\beta=\beta_c/0.8$ as obtained
         from $2 \times 2$ and $4 \times 4$ blocking on lattices
         with $L=32,\dots,512$. The quantity
         $\beta_{\rm eff}^{i,l}$ is defined in the text.
         For $L=256$ and $L=512$ we show results of two
         independent runs.}
\begin{center}
\begin{tabular}{|r|l|l|l|l|}
 \hline
     \multicolumn{1}{|c|}{$L$}
  &  \multicolumn{1}{c|}{$\beta_{\rm eff}^{1,2}$}
  &  \multicolumn{1}{c|}{$\beta_{\rm eff}^{1,4}$}
  &  \multicolumn{1}{c|}{$\beta_{\rm eff}^{2,2}$}
  &  \multicolumn{1}{c|}{$\beta_{\rm eff}^{2,4}$} \\
 \hline
 \hline
  32     & 4.50(8)  & 4.41(9)   &  4.46(3)   &  4.45(4)   \\
 \hline
  64     & 4.7(2)   & 4.74(20)  &  4.50(6)   &  4.48(7)   \\
 \hline
  128    & 4.04(18) & 4.03(21)  &  4.433(82) &  4.434(95)  \\
 \hline
  256    & 4.26(37) & 4.07(39)  &  4.37(17)  &  4.35(17)  \\
         & 3.9(3)   & 3.8(3)    &  4.30(15)  &  4.29(16)  \\
 \hline
  512    & 4.65(62) & 3.16(64)  &  5.02(35)  &  4.62(31)  \\
         & 5.1(6)   & 5.9(8)    &  4.25(21)  &  4.6(3)    \\
 \hline
 \end{tabular}
 \end{center}
 \end{table}
\end{document}